\documentclass[twocolumn,english,journal]{IEEEtran}
\usepackage[T1]{fontenc}
\usepackage[latin9]{inputenc}
\usepackage{units}
\usepackage{amsmath}
\usepackage{amssymb}
\usepackage{graphicx}

\makeatletter

\usepackage{tikz}
\usetikzlibrary{calc}

\@ifundefined{date}{}{\date{}}
\usepackage{cite}
\usepackage[printonlyused]{acronym}
\acrodef{AWGN}{additive white Gaussian noise}
\acrodef{ASE}{amplified spontaneous emission}
\acrodef{QAM}{quadrature amplitude modulation}
\acrodef{PAM}{pulse amplitude modulation}
\acrodef{SE}{spectral efficiency}
\acrodef{SNR}{signal to noise ratio}
\acrodef{TX}{transmitter}
\acrodef{RX}{receiver}
\acrodef{BER}{bit error rate}
\acrodef{SER}{symbol error rate}
\acrodef{NFT}{nonlinear Fourier transform}
\acrodef{BNFT}{backward NFT}
\acrodef{FNFT}{forward NFT}
\acrodef{I-FNFT}{incremental FNFT}
\acrodef{DF-FNFT}{decision-feedback FNFT}
\acrodef{DF-BNFT}{decision-feedback BNFT}
\acrodef{NFDM}{nonlinear frequency-division multiplexing}
\acrodef{OFDM}{orthogonal frequency-division multiplexing}
\acrodef{NIS}{nonlinear inverse synthesis}
\acrodef{DAC}{digital-to-analog converter}
\acrodef{ADC}{analog-to-digital converter}
\acrodef{GVD}{group velocity dispersion}
\acrodef{SMF}{single mode fiber}
\acrodef{NCG}{Nystrom conjugate gradient}
\acrodef{B2B}{back-to-back}
\acrodef{EDC}{electronic dispersion compensation}
\acrodef{MAP}{maximum a posteriori probability}

\usepackage[linesnumbered,ruled,vlined]{algorithm2e}
\usepackage{algorithmic}
\usepackage{microtype}

\makeatother

\usepackage{babel}
\begin{document}
\title{A New Twist on Low-Complexity Digital Backpropagation}
\author{Stella Civelli\thanks{This work was partially supported by the European Union - Next Generation
EU under the Italian National Recovery and Resilience Plan (NRRP),
Mission 4, Component 2, Investment 1.3, CUP J53C22003120001, CUP B53C22003970001,
partnership on \textquotedblleft Telecommunications of the Future\textquotedblright{}
(PE00000001 - program \textquotedblleft RESTART\textquotedblright ).}\thanks{S.~Civelli is with the Cnr-Istituto di Elettronica e di Ingegneria
dell\textquoteright Informazione e delle Telecomunicazioni (CNR-IEIIT),
Pisa, Italy, and with the Telecommunications, Computer Engineering,
and Photonics (TeCIP) Institute, Scuola Superiore Sant'Anna, Pisa,
Italy. D.~P.~Jana was with the Telecommunications, Computer Engineering,
and Photonics (TeCIP) Institute, Scuola Superiore Sant'Anna, Pisa,
Italy and currently is with the College of Optics and Photonics (CREOL),
University of Central Florida, Orlando, Florida 32816, USA. E.~Forestieri,
and M.~Secondini are with the Telecommunications, Computer Engineering,
and Photonics (TeCIP) Institute, Scuola Superiore Sant'Anna, Pisa,
Italy, and with the National Laboratory of Photonic Networks, CNIT,
Pisa, Italy. Email: stella.civelli@cnr.it. Part of this work was presented
at the European Conference on Optical Communication 2024 in Frankfurt
\cite{civelli2024CBESSFMECOC}.}, \IEEEmembership{Member, IEEE}, Debi Pada Jana, \IEEEmembership{Member, IEEE},
Enrico Forestieri, \IEEEmembership{Senior Member, IEEE}, and Marco
Secondini, \IEEEmembership{Senior Member, IEEE}}
\maketitle
\begin{abstract}
This work proposes a novel low-complexity digital backpropagation
(DBP) method, with the goal of optimizing the trade-off between backpropagation
accuracy and complexity. The method combines a split step Fourier
method (SSFM)-like structure with a simplified logarithmic perturbation
method to obtain a high accuracy with a small number of DBP steps.
Subband processing and asymmetric steps with optimized splitting ratio
are also employed to further reduce the number of steps required to
achieve a prescribed performance.

The first part of the manuscript is dedicated to the derivation of
a simplified logarithmic-perturbation model for the propagation of
a dual-polarization multiband signal in an optical fiber, which serves
as a theoretical background for the development of the proposed coupled-band
enhanced split step Fourier method (CB-ESSFM) and for the analytical
calculation of the model coefficients. Next, the manuscript presents
a digital signal processing algorithm for the implementation of DBP
based on a discrete-time version of the model and an overlap-and-save
processing strategy. Practical approaches for the optimization of
the coefficients used in the algorithm and of the splitting ratio
of the asymmetric steps are also discussed. A detailed analysis of
the computational complexity of the algorithm is also presented.

Finally, the performance and complexity of the proposed DBP method
are investigated through numerical simulations and compared to those
of other methods. In a five-channel 100\,GHz-spaced wavelength division
multiplexing system over a 15$\mathbf{\times}$80\,km single-mode-fiber
link, the proposed CB-ESSFM achieves a gain of about 1~dB over simple
dispersion compensation with only 15 steps (corresponding to 681 real
multiplications per 2D symbol), with an improvement of 0.9 dB over
conventional SSFM and almost 0.4~dB over our previously proposed
ESSFM. Significant gains and improvements are obtained also at lower
complexity. For instance, the gain reduces to a still significant
value of 0.34~dB when a single DBP step is employed, requiring just
75 real multiplications per 2D symbol. A similar analysis is performed
also for longer links, confirming the good performance of the proposed
method.
\end{abstract}


\section{Introduction\label{sec:Introduction}}

\IEEEPARstart{T}{he} continuous growth of global data traffic is
pushing optical networks to their limits \cite{Essiambre:JLT0210,Secondini2020_OpticalFiberTelecommunications_VII},
driving the investigation of several potential solutions \cite{winzer2015scaling,Roadmap2016}.
Among these, improving the performance of existing long-haul links
by modifying the digital signal processing (DSP) of the transmitter
and receiver is particularly attractive due its low cost, feasibility
and immediate applicability \cite[Ch. 10]{Roadmap2016}\cite{secondini2019JLT}.

A major limitation in current fiber-optic coherent systems is Kerr
nonlinearity \cite{agrawal2013_nonlinear_fiber_optics_5e}, which
degrades system performance as optical power increases \cite{bononi2020fiber}.
To mitigate nonlinear impairments in optical fibers, various DSP techniques
have been proposed, including digital back-propagation (DBP) \cite{essiambre2005fibre,ip2008compensation},
Volterra series equalization \cite{guiomar2013simplified}, and maximum-likelihood
sequence detection \cite{koike2012high,Mar:JLT14}. In particular,
DBP aims to reverse the effects of channel propagation by digitally
emulating the propagation of the signal through a fictitious fiber
link, equal to the actual transmission link but reversed in space
and with opposite propagation parameters. The most common implementation
of DBP is based on the split-step Fourier method (SSFM) \cite{Hasegawa73},
\cite[Sec. 2.4.1]{agrawal2013_nonlinear_fiber_optics_5e}, which decomposes
the propagation process into a series of simple independent linear
and nonlinear steps. However, to accurately implement DBP, a large
number of SSFM steps may be required, resulting in significant computational
complexity, which limits its practical deployment.

Several SSFM-based methods have been proposed to achieve a good trade-off
between performance and complexity. Filtered DBP accounts for correlations
among neighboring symbols in the generation of the nonlinear phase
rotation in each step \cite{du:OE2010,rafique2011compensation,Li:OFC11}.
The enhanced SSFM (ESSFM) reduces the number of required steps by
improving the accuracy of the nonlinear step, accounting for the interaction
between dispersion and intrachannel nonlinearity, practically generalizing
filtered DBP through the use of a finite impulse response (FIR) filter
with optimized coefficients \cite{Sec:ECOC14,secondini_PNET2016}.
The SSFM and its more efficient variants can be directly used also
for wavelength-division multiplexing (WDM) signals, as long as the
channels are jointly represented as a single optical field. In this
case, however, it might be computationally more efficient to represent
the WDM channels separately and account for their interaction through
a systems of coupled NLSEs \cite{leibrich2003efficient,mateo2010efficient}.
This idea is exploited in the coupled-channel ESSFM (CC-ESSFM), which
improves on ESSFM by accounting also for interchannel nonlinear effects
in WDM systems \cite{civelli2021coupled,civelli2021ISWCS2021}. A
similar approach can be used also for the propagation of a single
channel by digitally dividing it into subbands \cite{ip2011complexity}.
Recently, machine learning (ML) techniques have also been explored
for nonlinearity compensation, such as learned DBP \cite{hager2020physics,oliari2020revisiting},
subband-processing learned DBP \cite{hager2018wideband}, fully learned
perturbation-based nonlinearity compensation \cite{luo2023deep},
and nonlinearity mitigation based on carrier phase recovery \cite{neves2024ofc}.

In this work, we propose a novel single-channel DBP technique, referred
to as coupled-band ESSFM (CB-ESSFM), which combines the SSFM structure
with a logarithmic-perturbation (LP) model \cite{forestieri2005solving,secondini2012analytical,Secondini:JLT2013-AIR}
and subband processing to increase the accuracy of nonlinear steps
while keeping their complexity comparable to that of conventional
SSFM. The proposed propagation model constitutes a common theoretical
basis for the derivation of the ESSFM and CC-ESSFM methods, as well
as the CB-ESSFM method proposed in this work. The proposed method
depends on a set of coefficients that can be either obtained analytically
from the LP model or optimized numerically to maximize the performance.
Asymmetric steps with an optimized splitting ratio are also employed
to further improve the performance of the method. Numerical simulations
demonstrate that the proposed CB-ESSFM significantly enhances the
performance of state-of-the-art DBP methods while maintaining low
computational complexity.

The manuscript is organized as follows. Section \ref{sec:Theoreticalback}
presents the theoretical background of the proposed CB-ESSFM and an
analytical expression for its coefficients. Section \ref{sec:Implementation}
covers the practical implementation of the CB-ESSFM, detailing the
digital signal processing algorithm, discussing the optimization of
both the splitting ratio and the coefficients, and analyzing the computational
complexity. In Section \ref{sec:results}, we verify the accuracy
of the model and demonstrate the performance of the proposed DBP approach
across various scenarios using numerical simulations. Finally, Section
\ref{sec:conclusion} summarizes the conclusions.

\section{Theoretical Model\label{sec:Theoreticalback}}

In this Section, we present the theoretical background for the derivation
of the proposed DBP technique. In particular, we derive a fiber propagation
model based on the frequency resolved logarithmic perturbation (FRLP)
\cite{secondini2012analytical,Secondini:JLT2013-AIR}, starting from
a single-polarization single-band scenario, and then extending it
to the dual-polarization and multi-band cases. This model establishes
a general theoretical framework for deriving the enhanced SSFM (ESSFM)
and its variants \cite[Ch. 2]{Sec:ECOC14,secondini_PNET2016,civelli2021coupled,civelli2021ISWCS2021}---including
the DBP method that will be described in Section\,\ref{subsec:The-coupled-bands-ESSFM}---and
for evaluating the coefficients of the method.

\subsection{Single-Polarization Single-Band Model\label{subsec:single-pol_single-band}}

In the single-polarization single-band scenario, the propagation of
the optical signal is governed by the nonlinear Schr\"odinger equation
(NLSE) \cite[Ch. 2]{agrawal2013_nonlinear_fiber_optics_5e,Secondini2020_OpticalFiberTelecommunications_VII}
\begin{equation}
\frac{\partial u(z,t)}{\partial z}=j\frac{\beta_{2}}{2}\frac{\partial^{2}u(z,t)}{\partial t^{2}}-j\gamma Pg(z)|u(z,t)|^{2}u(z,t)\label{eq:NLSE}
\end{equation}
where $u(z,t)$ is the lowpass equivalent representation of the propagating
signal, normalized to have unit power at any distance $z$ in the
link; $\beta_{2}$ the fiber dispersion parameter; $\gamma$ the nonlinear
coefficient; $g(z)$ the power profile along the link (due to attenuation
and amplification) normalized to the reference power $P$ (typically
taken as the launch power), so that $\sqrt{Pg(z)}u(z,t)$ is the actual
lowpass equivalent representation of the optical signal at distance
$z$, with power $Pg(z)$.

\begin{figure}
\begin{centering}
\includegraphics[width=1\columnwidth]{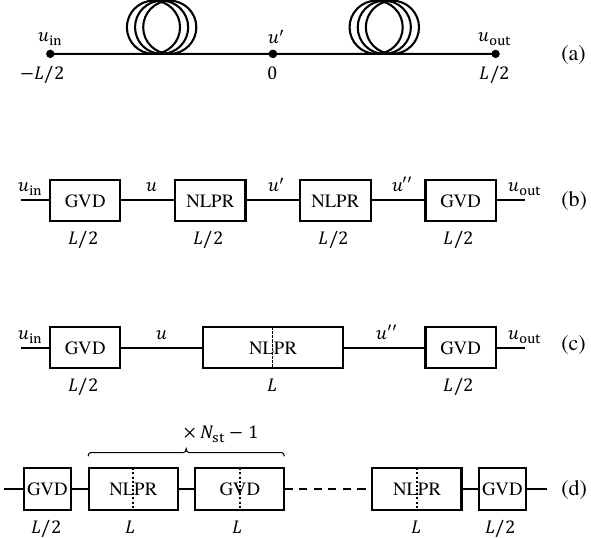}
\par\end{centering}
\caption{\label{fig:model_a_b_c_d}Derivation of the ESSFM model: (a) each
propagation step of length $L$ is divided into two halves; (b) the
approximated FRLP model is applied to each half (in reverse order
in the second half); (c) the two adjacent NLPR blocks are combined
into a single NLPR block; (d) the overall link of length $N_{\mathrm{st}}L$
is divided into $N_{\mathrm{st}}$ steps, each modeled as above, and
pairs of adjacent GVD blocks are combined into single GVD blocks.}
\end{figure}

As shown in Fig.~\ref{fig:model_a_b_c_d}(a), we consider the propagation
through a step of length $L$ and, without loss of generality, set
the origin $z=0$ in the middle of the step, dividing the propagation
into two half steps. The step represents a generic portion of the
link that possibly includes one or more spans of fibers (or fractions
of them) and optical amplifiers.
 We approximate the propagation from $u_{\mathrm{in}}(t)\triangleq u(-L/2,t)$
to $u'(t)\triangleq u(0,t)$ by applying an approximated version of
the FRLP model \cite{secondini2012analytical,Secondini:JLT2013-AIR},
which is derived in Appendix~\ref{app:model} and consists of a group-velocity
dispersion (GVD) block followed by a nonlinear phase rotation (NLPR).
On the other hand, using the same expedient employed in the symmetric
version of the classical SSFM \cite[Sec. 2.4.1]{nl-agrawal}, we approximate
the propagation from $u'(t)$ to $u_{\mathrm{out}}(t)\triangleq u(L/2,t)$
by applying the same model of the first half but in reverse order,
obtaining the block diagram in Fig.~\ref{fig:model_a_b_c_d}(b).
The cascade of the direct and reverse FRLP models ensures that error
terms with an odd symmetry around $z=0$ cancel out, resulting in
higher overall accuracy, and yields the symmetric configuration in
Fig.~\ref{fig:model_a_b_c_d}(c), where the two half-length NLPR
blocks are combined into a single NLPR block.\footnote{Strictly speaking, such a configuration is actually symmetric and
ensures error cancellation only when also the power profile $g(z)$
is symmetric around $z=0$, which is not true in general (e.g., for
long steps with relevant attenuation). In more general cases, a higher
accuracy is achieved by splitting the step asymmetrically, as discussed
in greater detail in Section~\ref{subsec:Optimization-of-the-splitting}.} After some calculation reported in Appendix~\ref{app:model}, we
obtain the following model\footnote{The Fourier transform of a generic function $x(t)$ is denoted with
the corresponding capital letter and defined as $X(f)=\int_{-\infty}^{+\infty}x(t)\exp\left(-j2\pi ft\right)\text{d}t$.
Also, note that, with an abuse of notation, we use the same symbol
$u$ to denote the propagating signal $u(z,t)$, the linearly propagated
signal after the first half step $u(t)$, and (later) its samples
$u[k]$. A similar notation will be used in the multi-band case.}
\begin{align}
u(t) & =\int_{-\infty}^{+\infty}U_{\mathrm{in}}(f)H(L/2,f)e^{j2\pi ft}df\label{eq:ESSFM-model_a}\\
u''(t) & =u(t)e^{-j\theta(t)}\label{eq:eq:ESSFM-model_b}\\
u_{\mathrm{out}}(t) & =\int_{-\infty}^{+\infty}U''(f)H(L/2,f)e^{j2\pi ft}df\label{eq:eq:ESSFM-model_c}
\end{align}
where
\begin{equation}
H(z,f)=\exp\left(-j2\pi^{2}\beta_{2}f^{2}z\right)\label{eq:GVD_transfer-function}
\end{equation}
is the transfer function due to the accumulated GVD after a propagation
distance $z$;
\begin{equation}
\theta(t)=P\iint_{\mathbb{R}^{2}}K(\mu,\nu)U(\mu)U^{*}(\nu)e^{j2\pi(\mu-\nu)t}d\mu d\nu\label{eq:NLPR_quadform}
\end{equation}
is the NLPR due to Kerr nonlinearity (and its interaction with GVD);
and
\begin{equation}
K(\mu,\nu)=\int_{-L/2}^{L/2}\gamma g(z)H(z,\mu)H^{*}(z,\nu)H(-z,\mu-\nu)dz\label{eq:kernel_definition}
\end{equation}
is a frequency-domain second-order Volterra kernel that accounts for
the efficiency with which pairs of frequency components of the propagating
signal contribute to the NLPR. A closed-form expression for (\ref{eq:kernel_definition})
is given in Section~\ref{subsec:Analytical-derivation-of}.

As in the classical SSFM, the propagation through a long link is usually
handled by dividing it into $N_{\mathrm{st}}$ steps, where $N_{\mathrm{st}}$
is selected to obtain the desired trade-off between complexity and
accuracy. When modeling each step as in Fig.~\ref{fig:model_a_b_c_d}(c),
pairs of adjacent half-length GVD blocks combine into single full-length
GVD blocks, yielding the overall block diagram in Fig.~\ref{fig:model_a_b_c_d}(d).

For a practical implementation of the model for the development of
DBP methods, which will be discussed in detail in Section~\ref{sec:Implementation},
it is convenient to consider a discrete-time representation of the
signals. As shown in Appendix~\ref{app:discrete-time}, by assuming
a sufficiently large sampling rate $R$,\footnote{In principle, the equation holds when the sampling rate is at least
equal to the Nyquist rate for $u(t)$, i.e., twice its bandwidth.
However, the NLPR induces some spectral broadening, meaning that an
accurate representation of the output signal $u_{\mathrm{out}}(t)$
may require a slightly larger sampling rate. This is important when
considering the cascade of many steps to represent the propagation
through a long link, as will be discussed in detail in Section~\ref{sec:Implementation}.} we obtain a discrete-time version of (\ref{eq:NLPR_quadform}), in
which the NLPR samples $\theta[k]\triangleq\theta(k/R)$ are related
to the time-domain samples of the linearly propagated signal $u[k]\triangleq u(k/R)$
through a discrete-time second-order Volterra kernel. The corresponding
expression is calculated in Appendix \ref{app:discrete-time} and,
as shown in (\ref{eq:theta_discrete_quadform}), still involves the
computation of an infinite number of terms. In practice, however,
the discrete-time second-order Volterra kernel, whose coefficients
$d[m,n]$ are defined in (\ref{eq:d_mn_coefficients}), has a limited
duration that depends mainly on the GVD accumulated over the step
length $L$ and (in minor part) on the pulse duration. This means
that it can be practically truncated to a finite number of terms.
Moreover, to further simplify the computation, we consider only the
diagonal terms of the kernel, represented by the coefficients $c[m]\triangleq d[m,m]$
and neglect the (usually smaller) off-diagonal terms $d[m,n]$, with
$m\neq n$. In this way, letting $N_{c}$ be the number of relevant
pre- and post-cursor coefficients, (\ref{eq:theta_discrete_quadform})
simplifies to
\begin{equation}
\theta[k]=\sum_{m=-N_{c}}^{N_{c}}c[m]|u[k-m]|^{2}\label{eq:NLPR_singleband_discrete}
\end{equation}
where
\begin{equation}
c[m]=\frac{P}{R^{2}}\int\limits _{-R/2}^{R/2}\int\limits _{-R/2}^{R/2}K(\mu,\nu)e^{j2\pi(\mu-\nu)m/R}d\mu d\nu\label{eq:ESSFM_coefficients_singleband}
\end{equation}
can be seen as the coefficients of the discrete-time impulse response
that relates the signal intensity to the NLPR. Equations (\ref{eq:ESSFM-model_a})--(\ref{eq:eq:ESSFM-model_c}),
(\ref{eq:NLPR_singleband_discrete}), and (\ref{eq:ESSFM_coefficients_singleband})
form the theoretical foundation for deriving the ESSFM and for analytically
computing its coefficients. The necessary model extensions to derive
the CC-ESSFM \cite{civelli2021coupled,civelli2021ISWCS2021} and the
CB-ESSFM are detailed in the upcoming sections.

\subsection{Single-Polarization Multiband Model\label{subsec:single-pol_multiband}}

The accuracy of the ESSFM model in Fig.~\ref{fig:model_a_b_c_d}(d)
depends on the amount of dispersion accumulated in each step, hence
on the number of steps $N_{\mathrm{st}}$ in which the link is divided.
The accumulated dispersion increases with the signal bandwidth, so
that wider bandwidth signals typically requires more steps to achieve
a prescribed accuracy. At the same time, more steps entails a higher
computational complexity for DBP implementation, as detailed in Section~\ref{subsec:Complexity}.
Therefore, a possible solution to reduce computational complexity
while maintaining high accuracy is to divide wide-band signals into
subbands.

The calculations carried out for the single-band case can be readily
extended to the multi-band scenario. In this context, the propagating
signal $u(z,t)$, with bandwidth $B$, is partitioned into $N_{\mathrm{sb}}$
subbands and expressed as
\begin{equation}
u(z,t)=\sum_{i=1}^{N_{\mathrm{sb}}}u_{i}(z,t)\label{eq:subband-division}
\end{equation}
where $u_{i}(z,t)$ is the signal portion contained in the $i$th
subband, centered at frequency $f_{i}$ and with a bandwidth $B/N_{\mathrm{sb}}$.\footnote{An alternative representation is often adopted, replacing each bandpass
component $u_{i}$ in (\ref{eq:subband-division}) with its equivalent
lowpass representation multiplied by the corresponding carrier frequency
term $\exp(j2\pi f_{i}t)$. This choice, however, entails a slightly
more complex formalism, as it requires the inclusion of some additional
terms in the set of coupled differential equations derived below to
account for phase shifts and walk-offs between subbands.}

Since the impact of dispersion within each subband is significantly
less pronounced than across the entire signal bandwidth, we can increase
the step size $L$ of the model (thereby using fewer steps for the
entire link) without sacrificing accuracy. At the same time, when
dividing the signal into subbands, we need to decide how to handle
the nonlinear interaction between them. The simplest option is to
neglect it, applying the ESSFM separately to each subband as if it
were the only propagating signal. In this way, we reduce the overall
complexity (thanks to the increased step size), but we also lose accuracy
(due to the neglected interband nonlinearity). An alternative option
is to replace (\ref{eq:subband-division}) in (\ref{eq:NLSE}), hence
expanding the nonlinear term in the latter into several sub-terms,
typically categorized as self-phase modulation (SPM), cross-phase
modulation (XPM), and four-wave mixing (FWM) terms \cite{agrawal2013_nonlinear_fiber_optics_5e}.
Taking into account all sub-terms in the model would significantly
amplify the complexity of each step, potentially nullifying the effort
made to reduce it. However, the various terms have varying degrees
of significance in contributing to the overall nonlinearity, particularly
in the presence of dispersion. The strategy, therefore, is to discard
less relevant terms, such as FWM ones, to strike the optimal balance
between accuracy and complexity.

By replacing (\ref{eq:subband-division}) in (\ref{eq:NLSE}), and
neglecting (non-degenerate) FWM terms, we obtain a set of $N_{\text{sb}}$
coupled NLSEs 
\begin{equation}
\begin{gathered}\frac{\partial u_{i}}{\partial z}=j\frac{\beta_{2}}{2}\frac{\partial^{2}u_{i}}{\partial t^{2}}-j\gamma Pg(z)\biggl(|u_{i}|^{2}+2\sum_{\ell\neq i}|u_{\ell}|^{2}\biggr)u_{i}\\
i=1,\ldots,N_{\mathrm{sb}}
\end{gathered}
\label{eq:coupled_NLSE_multiband}
\end{equation}
each including an SPM term and $N_{\mathrm{sb}}-1$ XPM terms.

As in the single-channel case, we approximate the propagation through
the step of length $L$ depicted in Fig.~\ref{fig:model_a_b_c_d}(a)
by using the FRLP model in a split-step symmetric configuration, including
this time also the XPM terms in the NLPR \cite[eq. (9)]{Secondini:JLT2013-AIR}.
The solution for the signal $u_{i}$ in the $i$th subband, whose
derivation is omitted due to its similarity with the single-band case,
can still be expressed as in (\ref{eq:ESSFM-model_a})--(\ref{eq:eq:ESSFM-model_c})
and visualized by the block diagram in Fig.~\ref{fig:model_a_b_c_d}(c)
(adding the subscript $i$ to all involved signals), but with a different
NLPR that includes both SPM and XPM terms 
\begin{align}
\theta_{i}(t) & =P\biggl(\iint_{\mathbb{R}^{2}}K(\mu,\nu)U_{i}(\mu)U_{i}^{*}(\nu)e^{2\pi j(\mu-\nu)t}d\mu d\nu\nonumber \\
 & \quad+\sum_{\ell\neq i}\iint_{\mathbb{R}^{2}}2K(\mu,\nu)U_{\ell}(\mu)U_{\ell}^{*}(\nu)e^{2\pi j(\mu-\nu)t}d\mu d\nu\biggr)\label{eq:NLPR_quadform_multiband}
\end{align}
Considering a discrete-time representation of the signals; letting
$u_{i}[k]\triangleq u_{i}(k/R')$ and $\theta_{i}[k]\triangleq\theta_{i}(k/R')$
be the samples of the signal and NLPR, respectively, taken at rate
$R'=R/N_{\mathrm{sb}}$; truncating the discrete-time Volterra kernel;
and neglecting its off-diagonal terms as in the single-band case,
we eventually obtain 
\begin{align}
\theta_{i}[k] & =\sum_{m=-N_{c}}^{N_{c}}c_{ii}[m]|u_{i}[k-m]|^{2}\nonumber \\
 & \quad+2\sum_{\ell\neq i}\sum_{m=-N_{c}}^{N_{c}}c_{i\ell}[m]|u_{\ell}[k-m]|^{2}\label{eq:NLPR_multiband_discrete}
\end{align}
where 
\begin{equation}
c_{i\ell}[m]=\frac{P}{R'^{2}}\int\limits _{f_{i\ell}-R'/2}^{f_{i\ell}+R'/2}\int\limits _{f_{i\ell}-R'/2}^{f_{i\ell}+R'/2}K(\mu,\nu)e^{j2\pi(\mu-\nu)m/R'}d\mu d\nu\label{eq:ESSFM_coefficients_general}
\end{equation}
is the $m$-th coefficient of the impulse response that accounts for
the impact of the intensity of the $\ell$th subband on the NLPR of
the $i$th subband, and $f_{i\ell}=f_{\ell}-f_{i}$ is the frequency
separation between the two subbands. For $i=\ell$, $f_{i\ell}=0$
and (\ref{eq:ESSFM_coefficients_general}) reduces to (\ref{eq:ESSFM_coefficients_singleband}).

\subsection{Dual-Polarization Multiband Model\label{subsec:dualpol_multiband}}

The previous model can be eventually extended to the case of a dual-polarization
signal, whose propagation is governed by the Manakov equation \cite{menyuk2006interaction}---formally
equal to (\ref{eq:NLSE}) but considering $u=(x,y)$ as the Jones
vector collecting the normalized complex envelopes of the two polarization
components, $x$ and $y$, and $|u|^{2}=|x|^{2}+|y|^{2}$ as the squared-norm
of the vector. As in (\ref{eq:subband-division}), we divide the signal
into $N_{\mathrm{sb}}$ subbands, where each subband contains two
polarization components $u_{i}=(x_{i},y_{i})$, and replace them in
(\ref{eq:NLSE}). In this case, besides SPM, XPM, and FWM, we obtain
also cross-polarization modulation (XPolM) terms \cite{Winter:JLT2009}.
Though XPolM can typically be as relevant as XPM, its effect cannot
be simply represented as an NLPR, and its inclusion in the model significantly
increases the complexity. In this case, we consider two possible solutions
to obtain a good trade-off between accuracy and complexity. The first
solution is obtained by omitting both FWM and XPolM terms, obtaining
a set of $N_{\text{sb}}$ coupled Manakov equations 
\begin{equation}
\begin{gathered}\frac{\partial u_{i}}{\partial z}=j\frac{\beta_{2}}{2}\frac{\partial^{2}u_{i}}{\partial t^{2}}-j\gamma Pg(z)\left(|u_{i}|^{2}+\frac{3}{2}\sum_{\ell\neq i}|u_{\ell}|^{2}\right)u_{i}\\
i=1,\ldots,N_{\mathrm{sb}}
\end{gathered}
\label{eq:coupled_Manakov_multiband}
\end{equation}
formally equivalent to (\ref{eq:coupled_NLSE_multiband}) but for
the 3/2 degeneracy factor in front of the XPM terms (instead of 2)
and for the vector interpretation as explained above. Consequently,
the propagation of the $i$th subband over the step of length $L$
depicted in Fig.~\ref{fig:model_a_b_c_d}(a) can still be expressed
as in (\ref{eq:ESSFM-model_a})--(\ref{eq:eq:ESSFM-model_c}) and
visualized by the block diagram in Fig.~\ref{fig:model_a_b_c_d}(c),
provided that the signals are interpreted as two-component vectors,
and the GVD and NLPR are applied to each polarization component. In
this case, considering a discrete-time representation of the signals,
the NLPR in the $i$th subband can be expressed as 
\begin{align}
\theta_{i}[k] & =\sum_{m=-N_{c}}^{N_{c}}c_{ii}[m]|u_{i}[k-m]|^{2}\nonumber \\
 & \quad+\frac{3}{2}\sum_{\ell\neq i}\sum_{m=-N_{c}}^{N_{c}}c_{i\ell}[m]|u_{\ell}[k-m]|^{2}\label{eq:NLPR-multiband-2pol}
\end{align}
i.e., as in the single-polarization case (\ref{eq:NLPR_multiband_discrete}),
but with the squared norm replacing the squared modulus, and a 3/2
degeneracy factor in front of the XPM terms. The coefficients are
the same as in (\ref{eq:ESSFM_coefficients_general}). This solution
provides the theoretical foundation for the reduced-complexity CC-ESSFM
in \cite{civelli2021ISWCS2021} and serves as a basis for developing
the CB-ESSFM in Section~\ref{subsec:The-coupled-bands-ESSFM}.

On the other hand, the second solution is obtained by neglecting
FWM terms and incorporating only certain XPolM terms in the multi-band
expansion of the vector form of (\ref{eq:NLSE}). Specifically, to
avoid a significant increase in computational complexity, we include
only those terms that can still be expressed as an NLPR, albeit of
a different entity for the two polarizations. In this case, it is
necessary to represent the evolution of the two polarization components
of each subband with two coupled equations, obtaining a set of $2N_{\mathrm{sb}}$
coupled differential equations \cite{secondini2019JLT} 
\begin{align}
\frac{\partial x_{i}}{\partial z} & =j\beta_{2}\frac{\partial^{2}x_{i}}{\partial t^{2}}-j\gamma Pg(z)\nonumber \\
 & \quad+\biggl(|x_{i}|^{2}+|y_{i}|^{2}+\sum_{\ell\neq i}2|x_{\ell}|^{2}+|y_{\ell}|^{2}\biggr)x_{i}\label{eq:coupled_NLSE_multibandl_polx}\\
\frac{\partial y_{i}}{\partial z} & =j\beta_{2}\frac{\partial^{2}y_{i}}{\partial t^{2}}-j\gamma Pg(z)\nonumber \\
 & \quad+\biggl(|x_{i}|^{2}+|y_{i}|^{2}+\sum_{\ell\neq i}|x_{\ell}|^{2}+2|y_{\ell}|^{2}\biggr)y_{i}\label{eq:coupled_NLSE_multibandl_poly}\\
 & i=1,\ldots,N_{\mathrm{sb}}\nonumber 
\end{align}
The equations above imply that the XPM induced by co-polarized subbands
is twice as effective as the XPM resulting from orthogonally-polarized
subbands. This stands in contrast to the simpler model in (\ref{eq:coupled_Manakov_multiband}),
where the same XPM affects both polarizations, corresponding to the
average XPM affecting the two polarizations in (\ref{eq:coupled_NLSE_multibandl_polx})
and (\ref{eq:coupled_NLSE_multibandl_poly}). For the rest, the equations
are similar to those in (\ref{eq:coupled_NLSE_multiband}), and the
application of the FRLP method, whose development is omitted due to
its similarity with the previous cases, yields a similar model. The
propagation of the $i$th subband over the step of length $L$ depicted
in Fig.~\ref{fig:model_a_b_c_d}(a) can still be expressed as in
(\ref{eq:ESSFM-model_a})--(\ref{eq:eq:ESSFM-model_c}), but replacing
the NLPR in (\ref{eq:eq:ESSFM-model_b}) with different NLPRs $\theta_{i}^{x}(t)$
and $\theta_{i}^{y}(t)$ for the two polarization components 
\begin{align}
x_{i}''(t) & =x_{i}(t)\exp\bigl(-j\theta_{i}^{x}(t)\bigr)\label{eq:model-polX}\\
y_{i}''(t) & =y_{i}(t)\exp\bigl(-j\theta_{i}^{y}(t)\bigr)\label{eq:model-polY}
\end{align}
Eventually, considering a discrete-time representation of the signals
and letting $x_{i}[k]=x_{i}(k/R')$, $y_{i}[k]=y_{i}(k/R')$, $\theta_{i}^{x}[k]=\theta_{i}^{x}(k/R')$,
and $\theta_{i}^{y}[k]=\theta_{i}^{y}(k/R')$, the NLPRs can be expressed
as 
\begin{align}
\theta_{i}^{x}[k] & =\sum_{m=-N_{c}}^{N_{c}}c_{ii}[m]\bigl(x_{i}^{2}[k+m]+y_{i}^{2}[k+m]\bigr)\nonumber \\
 & \quad+\sum_{\ell\neq i}\sum_{m=-N_{c}}^{N_{c}}c_{i\ell}[m]\bigl(2x_{\ell}^{2}[k+m]+y_{\ell}^{2}[k+m]\bigr)\!\label{eq:NLPR_polx}\\
\theta_{i}^{y}[k] & =\sum_{m=-N_{c}}^{N_{c}}c_{ii}[m]\bigl(x_{i}^{2}[k+m]+y_{i}^{2}[k+m]\bigr)\nonumber \\
 & \quad+\sum_{\ell\neq i}\sum_{m=-N_{c}}^{N_{c}}c_{i\ell}[m]\bigl(x_{\ell}^{2}[k+m]+2y_{\ell}^{2}[k+m]\bigr)\label{eq:NLPR_poly}
\end{align}
The coefficients $c_{i\ell}[m]$ are as in (\ref{eq:ESSFM_coefficients_general})
and do not depend on the particular polarization. This solution provides
the theoretical foundation for the full-complexity CC-ESSFM \cite{civelli2021coupled,civelli2021ISWCS2021}
and could be used also to develop a slightly different version of
the CB-ESSFM, though this is not explored in this work.

\subsection{Evaluation of the CB-ESSFM Coefficients\label{subsec:Analytical-derivation-of}}

In this Section, we provide an analytical expression for the kernel
function (\ref{eq:kernel_definition}) and a numerical procedure for
the evaluation of the CB-ESSFM coefficients defined in (\ref{eq:ESSFM_coefficients_general}).
For simplicity, we assume that the step in Fig.~\ref{fig:model_a_b_c_d}(a)
consists of $N_{\text{sp}}$ identical spans of fiber of length $L/N_{\text{sp}}$,
with attenuation coefficient $\alpha,$ dispersion parameter $\beta_{2}$,
nonlinear coefficient $\gamma$, and periodic amplification at the
end of each span which exactly compensates for the span loss. In this
case, letting $a=\alpha/2$ and $b\triangleq2\pi^{2}\beta_{2}\nu(\mu-\nu)$,
after a few calculations reported in Appendix~\ref{app:kernel},
the kernel function can be expressed as 
\begin{equation}
K(\mu,\nu)=\gamma e^{-aL/N_{\mathrm{sp}}}\frac{\sinh\bigl((a+jb)L/N_{\mathrm{sp}}\bigr)\sin(bL)}{(a+jb)\sin(bL/N_{\mathrm{sp}})}\label{eq:kernel_analytic}
\end{equation}
The expression (\ref{eq:kernel_analytic}) can be used also when the
step is just a fraction of a span. In this case, it is sufficient
to set $N_{\mathrm{sp}}=1$, $L$ to the actual length of the step
(not the whole span), and the reference power $P$ in (\ref{eq:ESSFM_coefficients_general})
to the power at the input of the step. For instance, if a span of
length $L_{\mathrm{sp}}$ is divided into three equal steps, the kernel
function is the same for all the steps and is obtained by setting
$L=L_{\mathrm{sp}}/3$ in (\ref{eq:kernel_analytic}), while the power
$P$ to be used in (\ref{eq:ESSFM_coefficients_general}) is different:
it equals the launch power in the first step, is attenuated by $\exp(-\alpha L)$
in the second step, and by $\exp(-2\alpha L)$ in the third step.
The approach can easily be extended to derive more general closed-form
expressions for steps that include several pieces of fiber with different
length and propagation parameters.

Given the kernel function (\ref{eq:kernel_analytic}), the coefficients
$c_{i\ell}[m]$ can be evaluated numerically from (\ref{eq:ESSFM_coefficients_general}),
e.g., as the diagonal terms of the two-dimensional FFT of the kernel
function (\ref{eq:kernel_analytic}), evaluated over a grid of $(2N_{c}+1)\times(2N_{c}+1)$
frequency values in the range $\mu,\nu\in[f_{i\ell}-R'/2,f_{i\ell}+R'/2]$.
Alternatively, as discussed in Section~\ref{sec:Implementation},
the CB-ESSFM coefficients can also be obtained through a numerical
optimization procedure aimed at minimizing/maximizing some loss/performance
metric (e.g., the mean square error). For a larger frequency separation
$f_{i\ell}$, the walk-off between subbands induced by dispersion
increases. Numerically, this effect results in a slower decay of the
magnitude of the coefficients $c_{i\ell}[m]$ with $m$. This implies
that, in general, the impulse response in (\ref{eq:NLPR_multiband_discrete})
can be truncated to a number of coefficients $N_{c}$ that depends
on $\ell-i$. This dependence, which we omitted in (\ref{eq:NLPR_multiband_discrete})
to keep the notation simple, will be analyzed and exploited in Section~\ref{subsec:Optimization-ESSFMcoeff}.

Here we report some general properties of the coefficients $c_{i\ell}[m]$,
which can be inferred from (\ref{eq:ESSFM_coefficients_general})
and (\ref{eq:kernel_analytic}) and can be used to simplify their
evaluation and optimization, as discussed in greater detail in Section~\ref{sec:Implementation}:
\begin{itemize}
\item they are linearly proportional to the product $\gamma P$;
\item they are independent of the modulation format (e.g., they are the
same for 16-QAM and 64-QAM, with and without shaping);
\item they depend on the frequency distance $f_{\ell}-f_{i}$ between the
interfering channel and the observed channel, not on their absolute
position, meaning that, for equally spaced channels, they depend on
$i$ and $\ell$ only through the difference $\ell-i$, so that
\begin{equation}
c_{i,\ell}[m]=c_{i+k,\ell+k}[m],\quad\forall k\in\mathbb{Z}\label{eq:frequency-shift-symmetry}
\end{equation}
\item they satisfy the symmetry condition
\begin{equation}
c_{i,\ell}[m]=c_{\ell,i}[-m]\label{eq:time-reversal-symmetry}
\end{equation}
implying that the SPM coefficients have always an even symmetry 
\begin{equation}
c_{i,i}[m]=c_{i,i}[-m].\label{eq:SPM_even_symmetry}
\end{equation}
\end{itemize}

\section{Implementation\label{sec:Implementation}}

This Section discusses the practical implementation of the proposed
DBP method and is divided into four parts. The first part describes
the signal processing algorithm. The second part explains the procedure
for the numerical optimization of the CB-ESSFM coefficients; the third
part introduces an alternative implementation based on asymmetric
steps and discusses the optimization of their splitting ratio. Finally,
the last part provides information about the computational complexity
of the method.

\subsection{The Coupled-Band ESSFM (CB-ESSFM)\label{subsec:The-coupled-bands-ESSFM}}

\begin{figure}
\begin{centering}
\includegraphics[width=1\columnwidth]{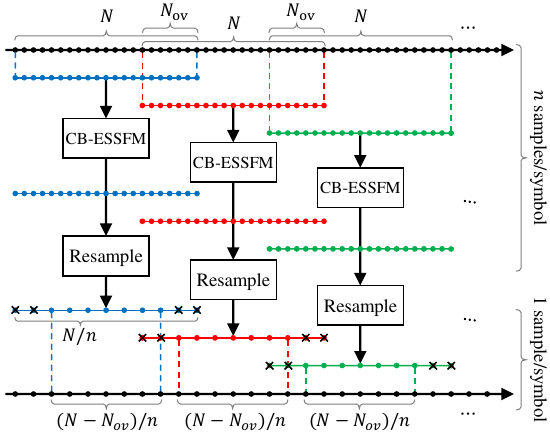}
\par\end{centering}
\caption{\label{fig:scheme_OeS}The received signal is processed by the CB-ESSFM
algorithm by using oversampling and overlap and save.}
\end{figure}

The proposed DBP algorithm is based on the theoretical model described
in the previous section and on the overlap-and-save technique \cite{oppenheim},
as schematically depicted in Fig.~\ref{fig:scheme_OeS}. First, the
received signal is sampled at $n$ samples per symbol, where the
oversampling factor $n\ge1+r$ is selected to account for the roll-off
factor $r$ of the modulation pulses and the bandwidth expansion induced
by nonlinearity. 
 The sequence of 4D samples (two polarizations and two quadratures
per sample) is then processed block-wise by using the overlap-and-save
technique \cite{oppenheim}. In practice, the sequence is divided
into partially overlapping blocks of $N$ samples, with an overlap
of $N_{\mathrm{ov}}$ samples. The overlap should at least equal the
duration of the overall channel response, which is mainly determined
by the accumulated GVD. Each block is processed by the CB-ESSFM algorithm,
obtaining the corresponding $N$ output samples, and then resampled
at one sample per symbol, obtaining $N/n$ samples. Eventually, the
overlapping samples are discarded---$N_{\mathrm{ov}}/2n$ on each
side of each block---while the remaining $(N-N_{\mathrm{ov}})/n$
per block are saved and recombined to form the whole output sequence.

\begin{figure}
\begin{centering}
\includegraphics[width=1\columnwidth]{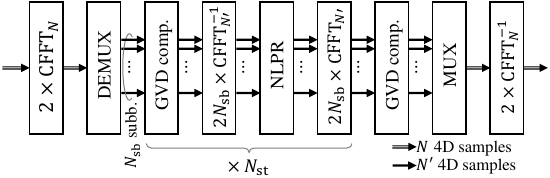}
\par\end{centering}
\caption{\label{fig:scheme_CB-ESSFM}CB-ESSFM algorithm with $N_{\mathrm{st}}$
steps and $N_{\mathrm{sb}}$ subbands.}
\end{figure}

\begin{figure}
\begin{centering}
\includegraphics[width=1\columnwidth]{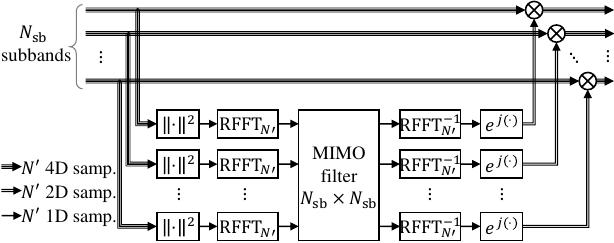}
\par\end{centering}
\caption{\label{fig:scheme_NLPR}Nonlinear phase rotation (NLPR) of the CB-ESSFM
algorithm.}
\end{figure}

The processing performed by the CB-ESSFM algorithm on each block of
$N$ 4D samples follows the scheme in Fig.~\ref{fig:model_a_b_c_d}(d),
with the dual-polarization multiband processing described in Section~\ref{subsec:dualpol_multiband}
and the NLPR in (\ref{eq:NLPR-multiband-2pol}).\footnote{We found that a full-complexity CB-ESSFM based on (\ref{eq:model-polX})--(\ref{eq:NLPR_poly})
only provides a minimal performance improvement in the considered
scenario. This improvement comes at a slightly higher computational
cost, without any practical benefits. Therefore, we decided not to
consider this implementation for the remainder of the paper.} The algorithm is further detailed in Fig.~\ref{fig:scheme_CB-ESSFM}
and \ref{fig:scheme_NLPR}.
At the input, two complex FFTs (CFFTs)---one for each polarization---are
performed on the whole block of $N$ samples, followed by a demultiplexer
(DEMUX) that divides the signal into $N_{\mathrm{sb}}$ subbands (at
no cost in the frequency domain), each represented by $N'=N/N_{\mathrm{sb}}$
samples. Then, GVD compensation and nonlinear phase rotation (NLPR)
are iteratively performed $N_{\mathrm{st}}$ times, each followed,
respectively, by $2N_{\mathrm{sb}}$ (one per each subband and polarization)
inverse and direct CFFTs of size $N'$ to perform GVD compensation
in the frequency domain and nonlinear phase rotation in the time domain.
Finally, one additional GVD compensation is performed, followed by
a multiplexer (MUX) that recombines the subbands and two inverse CFFTs
of size $N$.

The GVD compensation block consists in the multiplication of both
polarizations of each signal component (at frequency $f_{k}$) by
the corresponding value of the fiber transfer function $h_{k}=\exp(j2\pi^{2}\beta_{2}\Delta zf_{k}^{2})$,
where $\Delta z$ is the fiber length for which GVD compensation is
applied. In the configuration shown in Fig.~\ref{fig:model_a_b_c_d}(d),
the link is divided into $N_{\mathrm{st}}$ steps of length $L$,
and each step is symmetrically split into two halves of length $L/2$,
so that $\Delta z=L/2$ in the first and last GVD compensation blocks,
while $\Delta z=L$ in all the others. If necessary, the method could
be easily adjusted to accommodate a variable step size and a different
splitting ratio, with no significant changes to the implementation
scheme and complexity. The benefits of an asymmetric configuration
with an optimized splitting ratio are discussed in Section\ \ref{subsec:Optimization-of-the-splitting}.

The NLPR block follows the process shown in Fig.~\ref{fig:scheme_NLPR}.
 First, the intensity (squared norm) of the $N'$ 4D samples on each
subband is computed. Then, MIMO filtering of the $N_{\mathrm{sb}}$
real intensity signals is implemented in frequency domain to compute
the $N_{\mathrm{sb}}$ NLPRs. This operation, described in (\ref{eq:NLPR-multiband-2pol})
in the time domain, is equivalently but more efficiently implemented
in the frequency domain by performing $N_{\mathrm{sb}}$ real FFTs
(RFFTs) of size $N'$, multiplying the resulting vector of $N_{\mathrm{sb}}$
samples at each frequency component by the corresponding $N_{\mathrm{sb}}\times N_{\mathrm{sb}}$
MIMO transfer matrix, and then performing $N_{\mathrm{sb}}$ inverse
RFFTs of size $N'$ to go back to the time domain. The MIMO transfer
matrix is obtained offline by transforming (through FFT) the MIMO
impulse response matrix, whose elements correspond to the CB-ESSFM
coefficients in (\ref{eq:ESSFM_coefficients_general}). Finally, each
4D output sample is obtained by multiplying each polarization component
by the corresponding complex-exponential term (the same for both polarizations).

\begin{figure*}
\centering

\includegraphics[width=1.5\columnwidth]{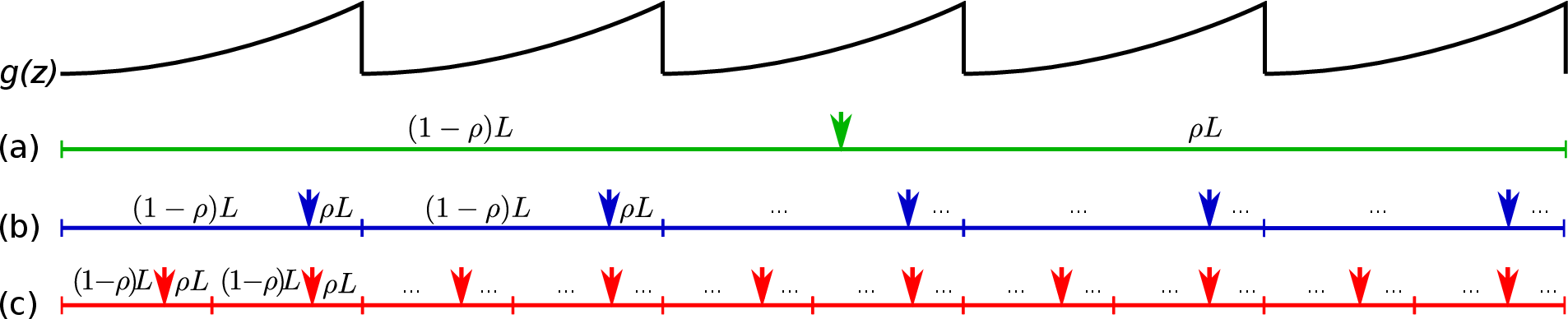}

\caption{\label{fig:ssfm-rhoopt}Power profile $g(z)$ in the DBP link and
three different step configurations: (a) 1 step/5 spans; (b) 1 step/span;
(c) 2 steps/span. The optimal positions of the NLPRs steps are denoted
by vertical arrows.}
\end{figure*}

\subsection{Optimization of the CB-ESSFM Coefficients\label{subsec:Optimization-ESSFMcoeff}}

The CB-ESSFM coefficients (or their FFT) needed for implementing the
MIMO filter in Fig.~\ref{fig:scheme_NLPR} can be obtained analytically
as explained in Section~\ref{subsec:Analytical-derivation-of}, with
minor modifications required to account for non-uniform span lengths
or asymmetric splitting ratios.

Alternatively, they can also be found by numerical (offline) optimization,
in order to maximize the system performance and/or consider unknown
or complex system configurations. In the remainder of the paper, unless
otherwise stated, we will use the latter approach, selecting the coefficients
that minimize the mean square error between the transmitted and received
symbols, using different data sets for optimization and performance
evaluation. To reduce the complexity of the optimization (which is,
however, done offline and does not affect the complexity of the DSP)
we
\begin{itemize}
\item [(i)] assume that the coefficients are the same in each step, but
for rescaling them proportionally to the signal power at the input
of each step when more than one step per span is used;
\item [(ii)] assume the symmetries in (\ref{eq:frequency-shift-symmetry})--(\ref{eq:SPM_even_symmetry});
\item [(iii)] consider a finite number of coefficients, which depends on
the channel memory: the number of coefficients for the interaction
of the subbands $\ell$ and $\ell+h$ is $2N_{c}(h)+1\approx\pi L\beta_{2}(nR/N_{\text{sb}})^{2}(h+1)$.
\end{itemize}
Given the above assumptions, letting $\mathbf{c}_{h}$ be the vector
that collects the $2N_{c}(h)+1$ coefficients that determine the interaction
between subbands $\ell$ and $\ell+h$ (the same for any $\ell$),
the optimization problem consists in finding the vectors $\mathbf{c}_{0},\mathbf{c}_{1},\ldots,\mathbf{c}_{N_{\mathrm{sb}}-1}$
that minimize the mean square error between the received symbols (after
DBP, matched filtering, resampling, and mean phase rotation removal)
and the transmitted symbols. To simplify the problem, we optimize
the $N_{\mathrm{sb}}$ vectors separately, in $N_{\mathrm{sb}}$ iterations:
in the $i$th iteration, we optimize $\mathbf{c}_{i-1}$, keeping
$\mathbf{c}_{0},\ldots,\mathbf{c}_{i-2}$ fixed at the values obtained
at the previous iterations, and setting $\mathbf{c}_{i},\ldots,\mathbf{c}_{N_{\mathrm{sb}}-1-2}$
to zero. The optimization of $\mathbf{c}_{i-1}$ is performed by using
Matlab's solver for nonlinear least squares problems based on the
trust-region-reflective algorithm, approximating the gradient through
finite differences. In our simulations, we did not observe substantial
convergence issues nor a critical dependence of the solution on the
initial values, which were simply set to the values corresponding
to a standard SSFM (all zeros, but for the central coefficient of
$\mathbf{c}_{0}$ equal to the average nonlinear phase rotation over
the step).

The assumptions made above could be relaxed to improve the performance
of the DBP algorithm. For instance, one could use a different set
of CB-ESSFM coefficients in each nonlinear step. However, this change
may lead to a more complex training phase. In this case, the use of
automatic differentiation tools for the optimization of the coefficients
might be beneficial, as they can efficiently compute gradients and
potentially enhance the convergence of the training process. This
possibility has not been investigated in this work and is left for
future study.

\subsection{Optimization of the Splitting Ratio \label{subsec:Optimization-of-the-splitting}}

In the symmetric configuration shown in Fig.\ \ref{fig:model_a_b_c_d}(a)--(d),
the NLPR is positioned in the middle of the step, sandwiched between
two equal GVD steps of length $L/2$. For a short piece of fiber with
negligible attenuation, this configuration ensures high accuracy thanks
to the cancellation of error terms with an odd symmetry around $z=0$.
This happens if the amount of nonlinearity and dispersion accumulated
in the two half steps is comparable. However, the ESSFM is designed
to reduce the number of propagation steps compared to the SSFM, with
each step possibly encompassing one or more fiber spans, so that the
power profile $g(z)$ (hence the accumulated nonlinearity) may be
highly asymmetric with respect to the middle of the step. In this
situation, an asymmetric configuration that divides the span in two
portions with comparable accumulated nonlinearity is expected to yield
better results. This even distribution of accumulated nonlinearity
is analogous to the criterion used in \cite{bosco2000suppression}
to derive the logarithmic step-size distribution for the SSFM.

Therefore, we consider a more general configuration in which the NLPR
is preceded by a GVD block of length $(1-\rho)L$ and followed by
a GVD block of length $\rho L$, where $\rho$ is referred to as splitting
ratio. As an example, Fig.~\ref{fig:ssfm-rhoopt} shows the power
profile $g(z)$ in a five-span (backward) link and, below, three different
DBP configurations: (a) a single step for the whole link; (b) one
step per span; and (c) two steps per span. The vertical arrows indicate
the NLPR position within each step. With a single step, a balanced
division of nonlinear effects is approximately obtained with a symmetric
configuration ($\rho=0.5$)---the approximation becoming more accurate
for a higher number of spans. For instance, a symmetric configuration
was shown to be optimal for the single-step DBP algorithm employed
in \cite{rafique2011compensation}. On the other hand, with one step
per span, nonlinear effects take place mostly in the last portion
of each step, so that an asymmetric configuration with $\rho<0.5$
results in a more balanced distribution of nonlinear effects in the
two portions of the step. Finally, with two steps per span, a balanced
distribution is still obtained with splitting ratio $\rho<0.5$, but
higher (approximately doubled) than in the previous case. Further
increasing the number of steps makes the attenuation in each step
less and less relevant, requiring the optimal $\rho$ to approach
again the symmetric configuration (as typically done in the conventional
SSFM). These heuristic considerations will be verified numerically
in Section~\ref{subsec:performance}.

With asymmetric steps, the overall DBP implementation changes only
slightly with respect to the symmetric case shown in Fig.~\ref{fig:model_a_b_c_d}(d).
In fact, couple of GVD blocks from adjacent steps can still be combined
to form a single GVD block of length $L$, and only the length of
the first and last GVD blocks must be modified to $(1-\rho)L$ and
$\rho L$, respectively.

\subsection{Complexity\label{subsec:Complexity}}

The computational complexity of the algorithm is assessed based on
the number of real multiplications (RMs) and real additions (RAs)
required for each 2D symbol. In our analysis, we assume that each
complex addition is carried out with two RAs, while each complex multiplication
(CM) requires three RMs and five RAs \cite{wenzler1995new}. If one
of the multipliers of a CM is known in advance, only three out of
the five RAs need to be computed in real-time, as two of them can
be precomputed offline. Analogously, if a couple of CMs involve the
same complex multiplier, two RAs can be computed only once and used
for both the CMs, so that each CM requires, on average, three RMs
and four RAs. Finally, if one of the multipliers is real, the CM reduces
to just two RMs. Moreover, we assume that the CFFT of $N$ complex
samples is implemented by the split-radix algorithm \cite{yavne1968economical},
which requires $N\log_{2}N-3N+4$ RMs and $3N\log_{2}N-3N+4$ RAs,
and that the RFFT requires approximately half the RMs and RAs of the
CFFT \cite{yavne1968economical}.\footnote{These assumptions differ from those in \cite{civelli2021ISWCS2021},
where we considered a naive CM implementation with four RMs and two
RAs, and the classical Cooley--Tukey FFT algorithm \cite{cooley1965algorithm}.
The present choice is slightly more efficient, particularly when RMs
are considered more expensive than RAs.}

At the output of the CB-ESSFM, each 4D output sample corresponds to
a couple of 2D symbols---one per polarization. Thus, the overall
computational complexity, is given by the following equations 
\begin{align}
C_{M} & =\frac{n\cdot N_{\mathrm{RM}}(N,N_{\mathrm{st}},N_{\mathrm{sb}})}{2\cdot(N-N_{\mathrm{ov}})} & \text{RM/2D symb.}\label{eq:cost_RM}\\
C_{A} & =\frac{n\cdot N_{\mathrm{RA}}(N,N_{\mathrm{st}},N_{\mathrm{sb}})}{2\cdot(N-N_{\mathrm{ov}})} & \text{RA/2D symb.}\label{eq:cost_RA}
\end{align}
where $N_{\mathrm{RM}}$ and $N_{\mathrm{RA}}$ denote, respectively,
the number of RMs and RAs required by the CB-ESSFM to process a block
of $N$ 4D samples, and we have made explicit their dependence on
$N$ and on the number of steps $N_{\mathrm{st}}$ and of subbands
$N_{\mathrm{sb}}$ employed by the algorithm.

The expression for $N_{\mathrm{RM}}$ and $N_{\mathrm{RA}}$ is derived
here. The CB-ESSFM in Fig.~\ref{fig:scheme_CB-ESSFM} applied to
a block of $N$ samples, with $N_{\text{sb}}$ bands, and $N_{\text{st}}$
steps uses $4$ (direct or inverse) CFFTs of size $N$, $4N_{\mathrm{sb}}N_{\mathrm{st}}$
CFFTs of size $N'=N/N_{\mathrm{sb}}$, $N_{\mathrm{st}}+1$ GVD compensations,
and $N_{\mathrm{st}}$ NLPRs. Concerning GVD compensation, it requires
only $2N$ CMs---two per each polarization of each 4D sample---with
a reduced complexity due to the fixed multiplier, and because the
factor $h_{k}$ is fixed and can be precomputed offline. This corresponds
to $6N$ RMs and $6N$ RAs for each GVD compensation step.

Regarding the cost of each NLPR, the computation of the intensity
of the $N'$ 4D samples uses $4N$ RMs and $3N$ RAs. For what concerns
the MIMO filtering, by exploiting the Hermitian symmetry of the $N'$
frequency components (both the signal intensity and the CB-ESSFM coefficients
are real in time domain), only half ($N'/2$) matrix--vector multiplications
need to be actually implemented. Considering that the elements of
the MIMO transfer matrix can be precomputed offline, and that the
diagonal elements are real due to the even symmetry of the SPM coefficients
in (\ref{eq:SPM_even_symmetry}), each matrix--vector multiplication
requires $N_{\mathrm{sb}}(N_{\mathrm{sb}}-1)$ CMs with one fixed
multiplier, $N_{\mathrm{sb}}$ CMs with one real multiplier, and $N_{\mathrm{sb}}(N_{\mathrm{sb}}-1)$
complex additions. Overall, MIMO filtering requires 
\begin{align*}
 & N\log_{2}N'+\frac{N}{2}(3N_{\mathrm{sb}}-7)+4N_{\mathrm{sb}} & \mathrm{RM}\\
 & 3N\log_{2}N'+\frac{N}{2}(5N_{\mathrm{sb}}-11)+4N_{\mathrm{sb}} & \mathrm{RA}
\end{align*}
For the computation of the complex exponential terms, we assume a
look-up-table implementation or similar approach, so we neglect its
complexity. The multiplication by the complex-exponential term requires
a couple of CMs with a common multiplier, resulting in $6N$ RMs and
$8N$ RAs to obtain the $N$ 4D samples ($N'$ on each subband). Overall,
by combining the cost of intensity computation, MIMO filtering, and
output CMs, the NLPR step in Fig.~\ref{fig:scheme_NLPR} requires
\begin{align*}
 & N\log_{2}N'+\frac{N}{2}(3N_{\mathrm{sb}}+13)+4N_{\mathrm{sb}} & \mathrm{RM}\\
 & 3N\log_{2}N'+\frac{N}{2}(5N_{\mathrm{sb}}+11)+4N_{\mathrm{sb}} & \mathrm{RA}
\end{align*}

In total, adding up the number of RMs and RAs required by all the
steps of the CB-ESSFM algorithm in Fig.~\ref{fig:scheme_OeS}, and
replacing them in (\ref{eq:cost_RM}) and (\ref{eq:cost_RA}), we
obtain the following equations for the computational complexity
\begin{align}
C_{M} & =\frac{n}{2}\frac{N}{N-N_{\mathrm{ov}}}\left((5N_{\mathrm{st}}+4)\log_{2}\frac{N}{N_{\mathrm{sb}}}+N_{\mathrm{st}}\frac{3N_{\mathrm{sb}}+1}{2}\right.\nonumber \\
 & \left.{}+4\log_{2}N_{\mathrm{sb}}-6+\frac{20N_{\mathrm{sb}}N_{\mathrm{st}}+16}{N}\right)\quad\mathrm{RM/2D\,symb.}\label{eq:costRM}
\end{align}
\begin{align}
C_{A} & =\frac{n}{2}\frac{N}{N-N_{\mathrm{ov}}}\left((15N_{\mathrm{st}}+12)\log_{2}\frac{N}{N_{\mathrm{sb}}}+N_{\mathrm{st}}\frac{5N_{\mathrm{sb}}-1}{2}\right.\nonumber \\
 & \left.{}+12\log_{2}N_{\mathrm{sb}}-6+\frac{20N_{\mathrm{sb}}N_{\mathrm{st}}+16}{N}\right)\quad\mathrm{RA/2D\,symb.}
\end{align}

For the sake of comparison, we report below the complexity of the
other DBP methods discussed in this work. The (single-band) ESSFM
can be obtained simply by considering a CB-ESSFM with a single band
(frequency-domain NLPR), or can be alternatively implemented with
a time-domain NLPR, with a complexity that depends on the number of
real symmetric CB-ESSFM coefficients $2N_{c}+1$ as
\begin{align}
C_{M} & =\frac{n}{2}\frac{N}{N-N_{\mathrm{ov}}}\left((N_{\mathrm{st}}+1)\left(4\log_{2}N-6+\frac{16}{N}\right)\right.\nonumber \\
 & \left.{}+N_{\mathrm{st}}(11+N_{c})\right)\quad\mathrm{RM/2D\,symb.}\label{eq:compl_RM_time}
\end{align}

\begin{align}
C_{A} & =\frac{n}{2}\frac{N}{N-N_{\mathrm{ov}}}\left((N_{\mathrm{st}}+1)\left(12\log_{2}N-6+\frac{16}{N}\right)\right.\nonumber \\
 & \left.{}+N_{\mathrm{st}}(11+2N_{c})\right)\quad\mathrm{RA/2D\,symb.}\label{eq:compl_RA_time}
\end{align}
The complexity of the SSFM and the optimized SSFM (OSSFM), corresponding
to a standard SSFM with optimized nonlinear coefficient, is obtained
by setting $N_{c}=0$ in (\ref{eq:compl_RM_time}) and (\ref{eq:compl_RA_time}),
while the complexity of GVD compensation is obtained by setting $N_{\mathrm{st}}=0$.

\section{Numerical results\label{sec:results}}

\subsection{System Description}

The link consists of $15$ (unless otherwise stated) spans of $80$\,km
single-mode fiber (SMF), with attenuation $\alpha_{\text{dB}}=\unit[0.2]{dB/km}$,
dispersion $D=17\thinspace\text{ps}/\text{nm}/\text{km}$, and Kerr
parameter $\gamma=1.27\thinspace\text{W}^{-1}\text{km}^{-1}$. After
each span, an erbium-doped fiber amplifier (EDFA) with a noise figure
of $4.5$\,dB compensates for the span loss. The transmitted signal
is made of five identical WDM channels with $100$\,GHz spacing.
Each channel carries a uniform dual-polarization 64-QAM signal, with
baud rate $R=93$\,GBd and (frequency-domain) root-raised-cosine
pulses with rolloff $r=0.05$. At the receiver side, the central channel
is demultiplexed and processed by either electronic dispersion compensation
(EDC) or digital backpropagation (DBP) with different possible algorithms:
the CB-ESSFM described in this paper; the ESSFM proposed in \cite{Sec:ECOC14,secondini_PNET2016},
equivalent to CB-ESSFM with $N_{\mathrm{sb}}=1$ and $\rho=0.5$;
the  OSSFM, corresponding to a standard SSFM with optimized nonlinear
coefficient and equivalent to an ESSFM with a single coefficient ($N_{c}=0$);
and ideal DBP, practically obtained from a standard SSFM with a very
large number of steps $N_{\mathrm{st}}$ ($N_{\mathrm{st}}\rightarrow\infty$),
sufficient to achieve performance saturation. In the overlap-and-save
algorithm, the number of overlapping samples is set to match the overall
channel memory ($N_{\mathrm{ov}}\approx1800$ for the $15\times\unit[80]{km}$
link), while the blocklength is selected as the power of two that
minimizes computational complexity ($N=16384$ for the $15\times\unit[80]{km}$
link).  All DBP algorithms use a uniform step size ($N_{\mathrm{st}}$
steps of the same length) and, unless otherwise stated, $n=1.125$
samples/symbol. Next, the signal undergoes matched filtering, resampling
at 1 sample/symbol, and mean phase rotation removal. Finally, the
performance is evaluated in terms of received signal-to-noise ratio
(SNR), where noise is defined as the difference between the output
samples and the transmitted symbols. The results shown in the following
are all obtained at the optimal launch power (approximately between
$3$ and $4$~dBm per channel).

\subsection{CB-ESSFM Coefficients: Analytical Evaluation vs Numerical Optimization\label{subsec:Comparison-coefficients}}

\begin{figure}
\begin{centering}
\centering\includegraphics[width=0.9\columnwidth]{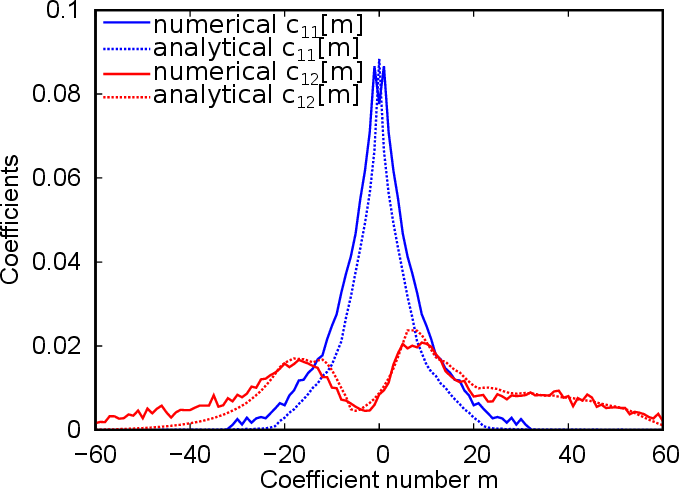}
\par\end{centering}
\begin{centering}
\par\end{centering}
\caption{\label{fig:Comparison-of-analytical}Comparison of intraband coefficients
$c_{11}[m]$ and interband coefficients $c_{12}[m]$ obtained analytically
or by numerical optimization for the case of $N_{\text{st}}=5$ steps.}
\end{figure}

Fig.~\ref{fig:Comparison-of-analytical} compares the CB-ESSFM coefficients
obtained analytically as explained in Section~\ref{subsec:Analytical-derivation-of}
(dashed) with those obtained by the numerical optimization procedure
described in Section~\ref{subsec:Optimization-ESSFMcoeff} (solid).
The coefficients are evaluated for the $15\times80$\,km link, considering
$N_{\text{st}}=5$ identical DBP steps (each made of three fiber spans)
and $N_{\text{sb}}=2$ bands, and their values are normalized to the
average nonlinear phase rotation over the step. The impulse response
determined by the intraband coefficients $c_{11}[m]$ (blue) is symmetric
and limited to approximately 50 samples. In contrast, the impulse
determined by the interband coefficients $c_{12}[m]$ (red) extends
to about 120 samples due to walk-off between the two bands, and exhibits
an irregular asymmetric shape caused by the combined effect of walk-off
and periodic attenuation/amplification. In both cases, there is good
agreement between the analytical and numerical curves, validating
the accuracy of the theoretical model developed in Section~\ref{sec:Theoreticalback}
and supporting the use of the analytical procedure for setting the
algorithm coefficients or as an initialization for numerical optimization.
In the next sections, we will always use the coefficients obtained
by numerical optimization.

\subsection{System Performance\label{subsec:performance}}

\begin{figure}
\centering\includegraphics[width=0.9\columnwidth]{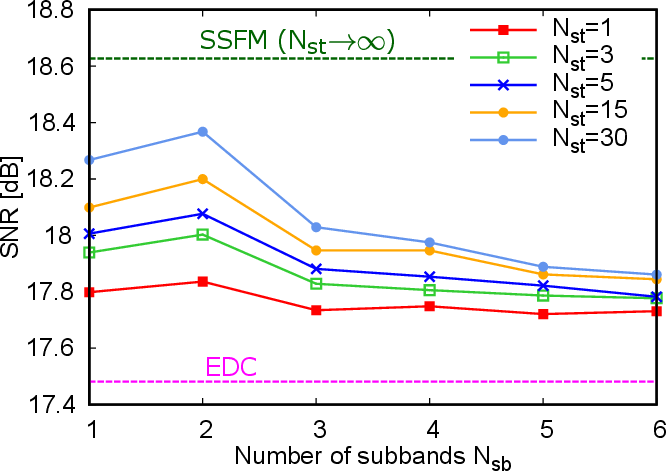}
\caption{\label{fig:SNR_vs_Nbands}SNR versus number of subbands $N_{\text{sb}}$
for the symmetric CB-ESSFM ($\rho=0.5)$ and various numbers of steps
$N_{\text{st}}$.}
\end{figure}

We start by investigating how the division in subbands and the optimization
of the splitting ratio affect the performance of the proposed algorithm.
Fig.~\ref{fig:SNR_vs_Nbands} shows the SNR obtained with the symmetric
CB-ESSFM ($\rho=0.5$) as a function of the number of subbands $N_{\text{sb}}$
for different numbers of steps $N_{\text{st}}$. As a reference, the
performance of EDC and ideal DBP are also shown. As expected, the
SNR improves when $N_{\text{st}}$ increases. On the other hand, as
discussed in Section~\ref{subsec:single-pol_multiband}, increasing
$N_{\mathrm{sb}}$ causes two opposite effects: the reduction of dispersion
within each subband, which improves model accuracy; and the appearance
of additional interband nonlinear interactions not included in the
model (FWM and XPolM), which degrades accuracy. In the current scenario,
regardless of the number of steps, the best tradeoff between the two
effects is obtained with $N_{\text{sb}}=2$ subbands, with a gain
of $0.1$\,dB with respect to the single-band case for $N_{\text{st}}=15$,
and a total gain of $\unit[0.72]{dB}$ with respect to EDC. On the
other hand, we expect that the optimal number of subbands depends
on the signal bandwidth and accumulated dispersion, meaning that more
subbands may be needed for higher baud rates, while a single band
might be optimal for low-dispersion links (e.g., in the O-band). The
investigation of different scenarios is left for future study.

\begin{figure}
\centering\includegraphics[width=0.9\columnwidth]{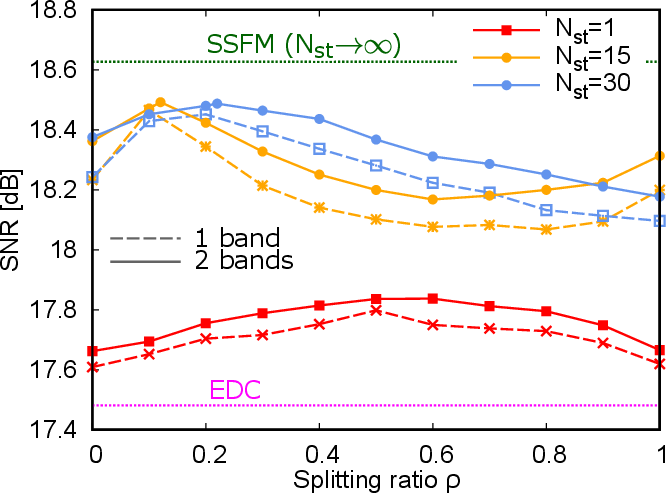}

\caption{\label{fig:SNR_vs_rho}SNR versus splitting ratio $\rho$ for the
CB-ESSFM with $N_{\text{sb}}=1$ band (dashed) and $N_{\text{sb}}=2$
bands (solid) and various numbers of steps $N_{\text{st}}$.}
\end{figure}

Next, Fig.\,\ref{fig:SNR_vs_rho} shows the SNR as a function of
the splitting ratio $\rho$ for different numbers of steps $N_{\mathrm{st}}$
and considering either a single band or two subbands. First, the figure
confirm that the two-subband implementation always outperforms the
single-band one, even when considering asymmetric steps with different
splitting ratios. Moreover, it shows that optimizing the splitting
ratio may be even more important in some cases. As conjectured in
Section~\ref{subsec:Optimization-of-the-splitting}, the best performance
is obtained when the step is split into two parts with comparable
nonlinear effects, resulting in more accurate error cancellation.
In fact, the symmetric configuration ($\rho=0.5$) is nearly optimal
when a single step is employed, but highly suboptimal for $N_{\mathrm{st}}=15$
(one step/span). In the latter case, the optimal splitting ratio is
$\rho\approx0.12$ (70~km/10~km splitting of the 80~km step) and
yields a gain of almost 0.3~dB with respect to the symmetric configuration,
with a total gain of about 1~dB compared to EDC. In this case, $\rho=0$
and $\rho=1$ are equivalent, as both place the nonlinear step at
the beginning of the span, resulting in identical performance. For
$N_{\mathrm{st}}=30$ (two steps/span), steps are shorter (40~km).
However, since nonlinear effects occur mostly toward the end of the
(backward) step, the optimal length $\rho L$ of the second part of
the step is only slightly reduced compared to the 80~km step, resulting
in an almost doubled optimal splitting ratio $\rho\approx0.22$ (31~km/9~km
splitting) and a lower gain of 0.12~dB compared to the symmetric
configuration. In this case, $\rho=0$ and $\rho=1$ result in different
nonlinear step configurations, with the former placing the nonlinear
step where the signal power is higher, thus providing better performance.
Finally, we note that with an optimized splitting ratio, increasing
the number of steps from $N_{\mathrm{st}}=15$ to $30$ does not provide
any improvement. We believe that this is due to two main reasons:
the performance approaching that of ideal DBP, and the negligible
amount of nonlinearity handled by odd steps compared to even steps
when using two equal steps per span. We expect that the performance
for $N_{\mathrm{st}}=30$ could be slightly improved by increasing
the length of odd steps compared to that of even steps. Indeed, optimizing
the step-size distribution may be beneficial in some scenarios (e.g.,
when multiple steps per span are used), particularly when combined
with the use of different CB-ESSFM coefficients in each step. This
additional optimization is left for future study.

\begin{figure}
\begin{centering}
\centering\includegraphics[width=0.9\columnwidth]{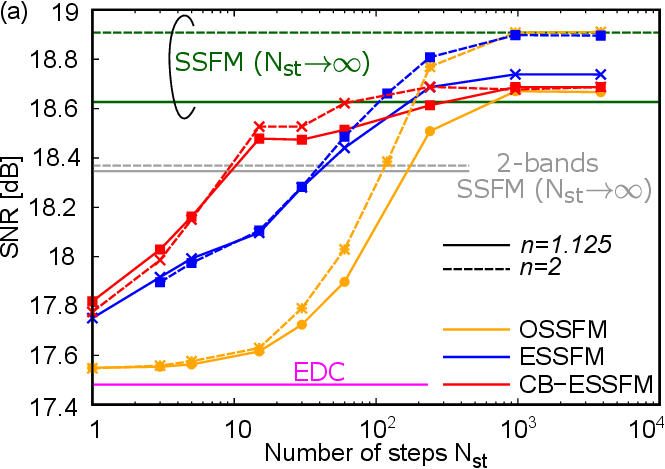}
\par\end{centering}
\begin{centering}
\includegraphics[width=0.9\columnwidth]{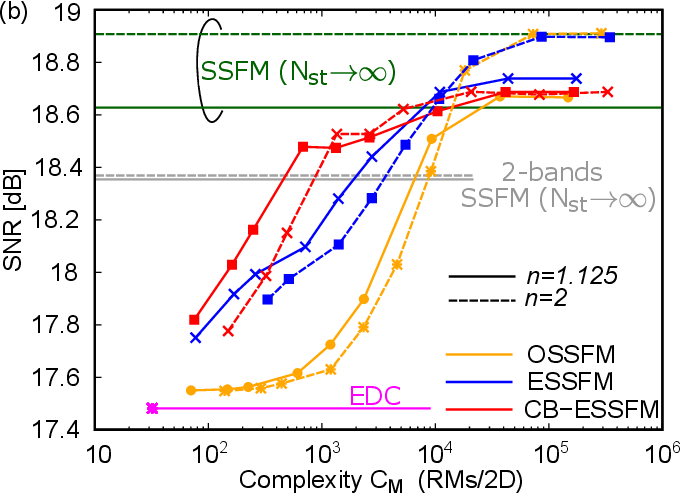}
\par\end{centering}
\caption{\label{fig:SNR_vs_compcompl}SNR for different DBP techniques and
oversampling factors versus (a) number of steps and (b) computational
complexity.}
\end{figure}

Next, we investigate the relation between performance and complexity
of the proposed algorithm, and how it compares to other DBP methods.
Based on the analysis above, all the CB-ESSFM results shown in the
following are obtained for $N_{\mathrm{sb}}=2$ and an optimized splitting
ratio. Fig.\,\ref{fig:SNR_vs_compcompl}(a) shows the performance
of different DBP techniques as a function of the number of steps $N_{\text{st}}$.
The solid curves are obtained using $n=1.125$ samples/symbol , while
the dashed ones are obtained with $n=2$ samples/symbol. Horizontal
lines indicating the performance of EDC and ideal DBP (with the two
oversampling factors, and for $1$ and $2$ bands) are also shown
as a reference. For a small number of steps, CB-ESSFM outperforms
all the other algorithms, approaching its best performance already
with $N_{\text{st}}=15$ steps. On the other hand, OSSFM gives almost
no gain up to 15 steps, needs hundreds of steps to achieve the performance
of CB-ESSFM, and saturates to its ultimate best performance for about
a thousand steps. Such large $N_{\mathrm{st}}$ values, though not
feasible for a practical implementation, give some additional insight
on the behavior of the algorithms. For instance, we see that OSSFM
and ESSFM saturate to a higher performance than CB-ESSFM, but only
if a significant oversampling ($n=2$) is employed. This is expected,
since the use of subband processing in CB-ESSFM entails neglecting
some interband nonlinear interactions, whose impact becomes relevant
only when all the other effects have been fully compensated by the
algorithm. In this case, also oversampling becomes important for OSSFM,
ESSFM, and the ideal single-band DBP curves when $N_{\text{st}}$
is large, but not for CB-ESSFM and two-bands DBP curves, which are
limited by interband nonlinearities. On the other hand, CB-ESSFM performs
significantly better than an ideal single-band DBP applied independently
to two subbands (shown with grey lines), confirming that the interband
XPM terms (neglected by the former) are accurately accounted for by
the coupled-band mechanism of CB-ESSFM.

\begin{figure}
\begin{centering}
\centering\includegraphics[width=0.9\columnwidth]{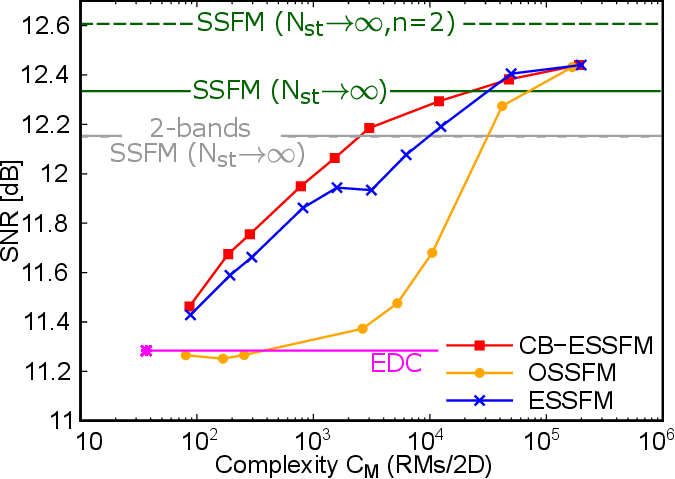}
\par\end{centering}
\caption{\label{fig:SNR_vs_compcompl-4800}SNR versus computational complexity
for the $\unit[60\times80]{km}$ link and different DBP techniques.}
\end{figure}

Next, Fig.\,\ref{fig:SNR_vs_compcompl}(b) reports the results in
Fig.\,\ref{fig:SNR_vs_compcompl}(a) as a function of the computational
complexity, defined as the number of required real multiplications
per 2D symbol and computed according to (\ref{eq:costRM}) and (\ref{eq:compl_RM_time}),
as discussed in Section~\ref{subsec:Complexity}. EDC is performed
in a single step and has a fixed complexity, represented by a single
purple square symbol in the figure. Additionally, a horizontal purple
line at the same SNR value is shown as a reference. Similarly, horizontal
lines indicate the SNR achievable by ideal DBP algorithms; their complexity
is considered infinite and, therefore, is not represented with symbols.
Fig.\,\ref{fig:SNR_vs_compcompl}(b) confirms and quantifies the
observations from Fig.\,\ref{fig:SNR_vs_compcompl}(a), showing that,
for reasonable levels of complexity ($<10^{3}$ RM/2D), the CB-ESSFM
with $n=1.125$ outperforms the other DBP techniques. Specifically,
the CB-ESSFM achieves a $0.34$dB gain at $\unit[75]{RM/2D}$, and
a 1~dB gain at $\unit[681]{RM/2D}$ compared to EDC, which requires
$\unit[32]{RM/2D}$. 
Fig.\,\ref{fig:SNR_vs_compcompl-4800} shows the performance versus
complexity for the same configuration as in Fig.\,\ref{fig:SNR_vs_compcompl}(b),
but with $60$ fiber spans, resulting in an overall link length of
$\unit[4800]{km}$. The observed behavior is similar to that of the
$\unit[1200]{km}$ link, though the required complexity is generally
higher; for instance, $10^{6}$ RM/2D are not sufficient to reach
saturation for both ESSFM and OSSFM. The largest gain of CB-ESSFM
over ESSFM is again achieved with one step per span, which in this
case corresponds to $N_{\text{st}}=60$ or approximately 3000 RM/2D.
We expect similar results even when a larger number of WDM channels
is used, possibly with a slightly smaller gain due to increased interchannel
nonlinearity. Conversely, when the bandwidth of each channel is larger,
the impact of intrachannel nonlinearity would be greater, allowing
higher gains, albeit with possibly increased complexity. In this case,
the optimal number of subbands may be larger.

\begin{figure}
\begin{centering}
\centering\includegraphics[width=0.9\columnwidth]{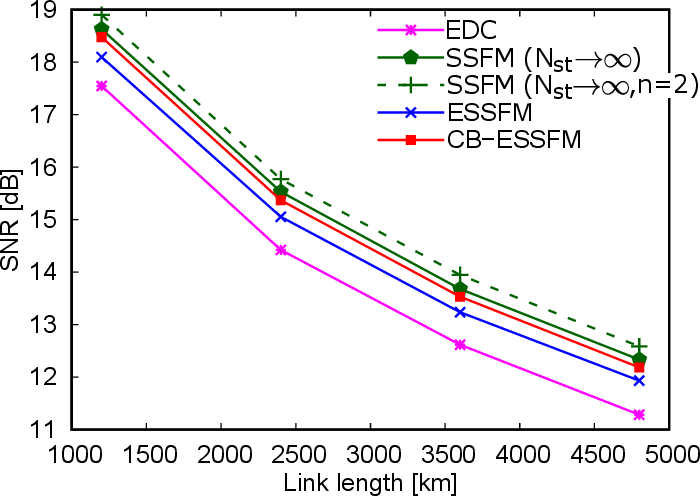}
\par\end{centering}
\caption{\label{fig:linklength}SNR versus link length for EDC and various
DBP techniques: ESSFM with $1$ step/span; CB-ESSFM with $1$ step/span,
$N_{\text{sb}}=2$ bands, and optimized $\rho$; ideal DBP (SSFM with
$N_{\mathrm{st}}\rightarrow\infty$) with different oversampling factors
($n=1.125$ except otherwise specified).}
\end{figure}

Finally, Fig.\,\ref{fig:linklength} compares the performance of
different DBP techniques as a function of link length, considering
$15$, $30$, $45$, and $60$ spans of 80km each. The figure shows
the SNR achieved with EDC, ideal DBP with $n=1.125$ or $n=2$ samples/symbol,
the symmetric ESSFM (i.e., with $\rho=0.5$), and the CB-ESSFM with
$N_{\text{sb}}=2$ subbands and an optimized splitting ratio. Both
ESSFM and CB-ESSFM are implemented with $1$ step per span (i.e.,
$15$, $30$, $45$, and $60$ steps for the respective lengths) and
$n=1.125$. The results indicate that the relative performance of
the methods and the gain provided by CB-ESSFM are consistently maintained
across different link lengths when using one step per span. 

\section{Conclusion\label{sec:conclusion}}

The proposed coupled-band enhanced split-step Fourier method (CB-ESSFM)
offers a promising approach to achieving a low-complexity, high-accuracy
trade-off in digital backpropagation (DBP) for fiber nonlinearity
mitigation. The approach combines an SSFM-like structure with a simplified
logarithmic perturbation to achieve a good accuracy with a reduced
number of steps. The use of subband processing and asymmetric steps
with an optimized splitting ratio enables a further reduction in the
required number of steps to achieve a desired performance.

The theoretical foundation for CB-ESSFM was developed by deriving
a simplified logarithmic-perturbation model for dual-polarization
multiband signals in optical fibers. This model provides a basis for
both the CB-ESSFM design and for the analytical computation of its
coefficients. Subsequently, a digital signal processing algorithm
was devised, based on a discrete-time version of the model and an
overlap-and-save strategy. Practical methods for optimizing the algorithm\textquoteright s
coefficients and the asymmetric step splitting ratio were introduced,
with a comprehensive analysis of the computational complexity.

Through numerical simulations, the performance and complexity of CB-ESSFM
were shown to be superior to existing DBP methods. In a 5$\times$93~GBd
WDM system over a 15-span (80 km each) single-mode fiber link, a 15-step
CB-ESSFM, requiring 681 real multiplications per 2D symbol (RM/2D),
demonstrated a 1~dB gain over simple dispersion compensation, with
an improvement (for the same complexity) of 0.4~dB compared to the
previously proposed ESSFM method, and 0.9~dB compared to a conventional
optimized SSFM. Notable gains were also achieved at lower complexity
levels; for example, gains of 0.34 dB, 0.55~dB, and 0.7~dB were
obtained with 75, 162, and 248 RM/2D, respectively.

The performance trends observed for shorter links were consistent
in longer links, though with a nearly proportional increase of complexity,
confirming the robustness of CB-ESSFM across varying transmission
distances. These results highlight the potential of CB-ESSFM as an
efficient DBP approach for fiber nonlinearity compensation, making
it an attractive candidate for practical deployment in long-haul optical
networks where minimizing complexity without compromising performance
is essential. Future work to further highlight and enhance the potential
of CB-ESSFM include (i) the investigation of different scenarios (e.g.,
higher baud rates, single-span links, O-band); (ii) the joint optimization
of the step-size distribution, splitting ratio, and of different CB-ESSFM
coefficients for each step; and (iii) the use of automatic differentiation
tools to enhance the convergence of the training process for such
more complex optimization problems.

\appendices{}

\section{{}\label{app:model}}

According to the FRLP model, the propagation through the first half
of the step in Fig.~\ref{fig:model_a_b_c_d}(a) can be approximated
as \cite[eq. (7)]{Secondini:JLT2013-AIR}, 
\begin{align}
u'(t) & \approx\int_{-\infty}^{+\infty}U_{\mathrm{in}}(f)H(L/2,f)e^{-j\theta'(f,t)}e^{j2\pi ft}df\label{eq:FRLP_original_model}
\end{align}
where $U_{\mathrm{in}}(f)$ is the Fourier transform of $u_{\mathrm{in}}(t)$,
$H(z,f)$ is given in (\ref{eq:GVD_transfer-function}), and $\theta'(f,t)$
is a frequency- and time-dependent nonlinear phase rotation (NLPR)
that accounts for SPM and its interaction with dispersion and attenuation
\cite[eq. (8)]{Secondini:JLT2013-AIR}. To further simplify the model,
we neglect the frequency dependence of the NLPR and replace it with
its value in the middle of the signal bandwidth, $\theta'(t)\triangleq\theta'(0,t)$.\footnote{A more accurate approximation could be obtained by replacing $\theta'(f,t)$
with its average or effective value over the signal bandwidth.} By letting $f=0$ in \cite[eq. (8)]{Secondini:JLT2013-AIR}, representing
the input signal $U_{\mathrm{in}}(f)$ as a function of the linearly
propagated signal $U(f)=U_{\mathrm{in}}(f)H(L/2,f)$, the NLPR can
be expressed as
\begin{equation}
\theta'(t)=P\iint_{\mathbb{R}^{2}}K'(\mu,\nu)U(\mu)U^{*}(\nu)e^{j2\pi(\mu-\nu)t}\text{d}\mu\text{d}\nu\label{eq:NLPR_1st_half}
\end{equation}
where
\begin{equation}
K'(\mu,\nu)=\int_{-L/2}^{0}\gamma g(z)H(z,\mu)H^{*}(z,\nu)H^{*}(z,\mu-\nu)dz\label{eq:kernel_1st_half}
\end{equation}
is the kernel function, properly redefined with respect to \cite[eq. (10)]{Secondini:JLT2013-AIR}
to account for the additional approximation ($f=0$) and different
normalizations used here. With this approximation, (\ref{eq:FRLP_original_model})
can be rewritten as
\begin{align}
u(t) & =\int_{-\infty}^{+\infty}U_{\mathrm{in}}(f)H(L/2,f)e^{j2\pi ft}df\\
u'(t) & =e^{-j\theta'(t)}u(t)
\end{align}
which can be interpreted as the cascade of the GVD and NLPR accumulated
over the first half of the step, and represented by the first two
blocks in Fig.~\ref{fig:model_a_b_c_d}(b). The propagation through
the second half of the step in Fig.~\ref{fig:model_a_b_c_d}(a) is
then approximated by a reverse-order FRLP model (with the NLPR preceding
the GVD). This is obtained by using the same approximated FRLP model
derived above to describe the backward propagation from $L/2$ to
0, obtaining
\begin{align}
u''(t) & =\int_{-\infty}^{+\infty}U_{\mathrm{out}}(f)H(-L/2,f)e^{j2\pi ft}df\label{eq:FRLP_model_reverse_a}\\
u'(t) & =e^{-j\theta''(t)}u''(t)\label{eq:FRLP_model_reverse_b}
\end{align}
where the NLPR is
\begin{align}
\theta''(t) & =P\iint_{\mathbb{R}^{2}}K''(\mu,\nu)U''(\mu)U''^{*}(\nu)e^{j2\pi(\mu-\nu)t}\text{d}\mu\text{d}\nu\nonumber \\
 & \approx P\iint_{\mathbb{R}^{2}}K''(\mu,\nu)U(\mu)U^{*}(\nu)e^{j2\pi(\mu-\nu)t}\text{d}\mu\text{d}\nu\label{eq:NLPR_2nd_half}
\end{align}
with kernel function
\begin{equation}
K''(\mu,\nu)=\int_{L/2}^{0}\gamma g(z)H(z,\mu)H^{*}(z,\nu)H^{*}(z,\mu-\nu)dz\label{eq:Ksinglecha-2-1}
\end{equation}
The approximation in (\ref{eq:NLPR_2nd_half}) consists in replacing
the potential $|u|^{2}$ that appears in the nonlinear term of the
NLSE (\ref{eq:NLSE}) and drives the generation of the NLPR with the
intensity of the linearly propagated signal. This is the same approximation
employed in \cite{secondini2012analytical,Secondini:JLT2013-AIR}
to derive the original FRLP model in (\ref{eq:FRLP_original_model}),
so it is not expected to further reduce the accuracy of the final
model. The forward propagation model for the second half of the step
is obtained by inverting (\ref{eq:FRLP_model_reverse_a}) and (\ref{eq:FRLP_model_reverse_b}),
obtaining
\begin{align}
u''(t) & =e^{j\theta''(t)}u'(t)\\
u_{\mathrm{out}}(t) & =\int_{-\infty}^{+\infty}U''(f)H(L/2,f)e^{j2\pi ft}df
\end{align}
corresponding, respectively, to the second NLPR and GVD blocks in
Fig.~\ref{fig:model_a_b_c_d}(b). The model can be further simplified
by combining the two adjacent NLPR blocks in Fig.~\ref{fig:model_a_b_c_d}(b)
into a single block with overall NLPR $\theta(t)=\theta'(t)-\theta''(t)$,
as shown in Fig.~\ref{fig:model_a_b_c_d}(c), resulting in the overall
propagation model (\ref{eq:ESSFM-model_a})--(\ref{eq:eq:ESSFM-model_c}),
with kernel function $K(\mu,\nu)=K'(\mu,\nu)-K''(\mu,\nu)$ given
by (\ref{eq:kernel_definition}).

\section{{}\label{app:discrete-time}}

We assume that $u(t)$ is band-limited, so that it can be represented
as
\begin{equation}
u(t)=\sum_{k=-\infty}^{\infty}u[k]\text{sinc}\left(Rt-k\right)\label{eq:u(t)_sampled}
\end{equation}
where $u[k]=u(k/R)$ are its samples taken at sufficiently high rate
$R=1/T$. By taking the Fourier transform of (\ref{eq:u(t)_sampled})
\begin{equation}
U(f)=\frac{1}{R}\sum_{k=-\infty}^{\infty}u[k]\text{rect}\left(f/R\right)\exp(-j2\pi kf/R)
\end{equation}
and replacing it in (\ref{eq:NLPR_quadform}), we can express the
NLPR samples as 
\begin{align}
\theta[k] & =\theta(k/R)\nonumber \\
 & =(P/R^{2})\sum_{m=-\infty}^{\infty}\sum_{n=-\infty}^{\infty}u[m]u^{*}[n]\nonumber \\
 & \times\int\limits _{-R/2}^{R/2}\int\limits _{R/2}^{R/2}K(\mu,\nu)e^{j2\pi[(k-m)\mu-(k-n)\nu)/R}d\mu d\nu\nonumber \\
 & =\sum_{m=-\infty}^{\infty}\sum_{n=-\infty}^{\infty}d[k-m,k-n]u[m]u^{*}[n]\nonumber \\
 & =\sum_{m=-\infty}^{\infty}\sum_{n=-\infty}^{\infty}d[m,n]u[k-m]u^{*}[k-n]\label{eq:theta_discrete_quadform}
\end{align}
where 
\begin{equation}
d[m,n]=\frac{P}{R^{2}}\int\limits _{-R/2}^{R/2}\int\limits _{-R/2}^{R/2}K(\mu,\nu)e^{j2\pi(m\mu-n\nu)/R}d\mu d\nu\label{eq:d_mn_coefficients}
\end{equation}
are the coefficients of the discrete-time second-order Volterra kernel,
corresponding to the samples of the inverse two-dimensional Fourier
transform of the frequency-domain Volterra kernel (\ref{eq:kernel_definition})
over a bandwidth equal to the sampling rate. After truncating the
discrete-time Volterra kernel and neglecting its off-diagonal terms
as discussed in Section~\ref{subsec:single-pol_single-band}, we
finally obtain the NLPR expression in (\ref{eq:NLPR_singleband_discrete})
with the CB-ESSFM coefficients in (\ref{eq:ESSFM_coefficients_singleband}).

In the multiband case, the derivation of the discrete-time model proceeds
analogously. After dividing the signal $u(t)$ into $N_{\mathrm{sb}}$
subbands as in (\ref{eq:subband-division}), the generic $i$th subband
is represented as 
\begin{align}
u_{i}(t) & =\sum_{k=-\infty}^{\infty}u_{i}[k]\text{sinc}\left(R't-k\right)\exp\left(j2\pi f_{i}t\right)\label{eq:subband_sampled}
\end{align}
where $u_{i}[k]=u(k/R')\exp(-j2\pi f_{i}k/R')$ is the $k$th sample
of the lowpass equivalent representation of the signal in the $i$th
subband, taken at rate $R'=R/N_{\mathrm{sb}}$. By replacing the Fourier
transform of (\ref{eq:subband_sampled}) in (\ref{eq:NLPR_quadform_multiband}),
truncating the discrete-time Volterra kernel and neglecting its off-diagonal
terms as in the single-band case, we finally obtain the NLPR expression
in (\ref{eq:NLPR_multiband_discrete}) with the CB-ESSFM coefficients
in (\ref{eq:ESSFM_coefficients_general}). The derivations of the
dual-polarization discrete-time models (\ref{eq:NLPR-multiband-2pol})
or (\ref{eq:NLPR_polx})--(\ref{eq:NLPR_poly}) are very similar
and are omitted.

\section{{}\label{app:kernel}}

By replacing (\ref{eq:GVD_transfer-function}) in (\ref{eq:kernel_definition}),
and defining
\begin{equation}
b\triangleq2\pi^{2}\beta_{2}\nu(\mu-\nu)
\end{equation}
the kernel function can be rewritten as 
\begin{equation}
K(\mu,\nu)=\int_{-L/2}^{L/2}\gamma g(z)\exp(-j2bz)dz\label{eq:kernel_b}
\end{equation}
For $N_{\mathrm{sp}}$ even, letting $\zeta=z-\left\lfloor z/L_{\mathrm{sp}}\right\rfloor \cdot L_{\mathrm{sp}}$,
so that $g(z)=e^{-\alpha\zeta}$, we can rewrite (\ref{eq:kernel_b})
as
\begin{alignat}{1}
 & \sum_{n=-N_{\mathrm{sp}}/2}^{N_{\mathrm{sp}}/2-1}\exp(-j2bnL_{\mathrm{sp}})\int_{0}^{L_{\mathrm{sp}}}\gamma\exp[-(\alpha+j2b)\zeta]d\zeta\nonumber \\
 & =\gamma\frac{1-e^{-(\alpha+j2b)L_{\mathrm{sp}}}}{\alpha+j2b}\sum_{n=-N_{\mathrm{sp}}/2}^{N_{\mathrm{sp}}/2-1}\exp(-j2bnL_{\mathrm{sp}})\nonumber \\
 & =\gamma\frac{1-e^{-(\alpha+j2b)L_{\mathrm{sp}}}}{\alpha+j2b}\exp(jbN_{\mathrm{sp}}L_{\mathrm{sp}})\frac{1-\exp(-j2bN_{\mathrm{sp}}L_{\mathrm{sp}})}{1-\exp(-j2bL_{\mathrm{sp}})}\label{eq:kernel_numerator}
\end{alignat}

Eventually, after a few passages, we obtain the expression in (\ref{eq:kernel_analytic})
for the kernel function. The same result is obtained, with similar
calculations, when $N_{\mathrm{sp}}$ is odd.

\bibliographystyle{ieeetr}

\end{document}